\documentclass[10pt,conference]{IEEEtran}
\IEEEoverridecommandlockouts
\usepackage{cite}
\usepackage{amsmath,amssymb,amsfonts,amsthm, empheq}
\usepackage{algorithmic}
\usepackage{graphicx}
\usepackage{textcomp}
\usepackage{xcolor}

\usepackage[english]{babel}
\usepackage{acronym}
\usepackage{hyperref}
\usepackage{booktabs}
\usepackage{caption}
\usepackage{subcaption}
\usepackage{float}

\renewcommand{\thetable}{\arabic{table}}

\newcommand{\eq}[1]{Eq.~\eqref{#1}}

\newcommand{\fig}[1]{Fig.~\ref{#1}}
\newcommand{\tab}[1]{Tab.~\ref{#1}}
\newcommand{\secref}[1]{Section~\ref{#1}}

\newcommand{\vv}[1]{\mathbf{#1}}

\graphicspath{{./figures/}}

\acrodef{prop}[\textit{MIMORPH}]{MIMO Radio Platform for Heterogeneous wireless systems}

\acrodef{3gpp}[3GPP]{3rd Generation Partnership Project}
\acrodef{abft}[A-BFT]{Association Beamforming Training}
\acrodef{ack}[ACK]{Acknowledgment}
\acrodef{adc}[ADC]{Analog-to-Digital Converter}
\acrodef{aoa}[AoA]{Angle of Arrival}
\acrodef{aod}[AoD]{Angle of Departure}
\acrodef{ap}[AP]{Access Point}
\acrodef{amc}[AMC]{Advanced Mezzanine Card}
\acrodef{awv}[AWV]{Antenna Wave Vector}
\acrodef{axi}[AXI]{Advanced eXtensible Interface}
\acrodef{ber}[BER]{Bit Error Rate}
\acrodef{bft}[BFT]{Beamforming Training}
\acrodef{bp}[BP]{Beam Pattern}
\acrodef{brp}[BRP]{Beam Refinement Phase}
\acrodef{cs}[CS]{Compressed Sensing}
\acrodef{cdf}[CDF]{Cumulative Distribution Function}
\acrodef{cef}[CEF]{Channel Estimation Field}
\acrodef{cfo}[CFO]{Carrier Frequency Offset}
\acrodef{sfo}[SFO]{Sampling Frequency Offset}
\acrodef{cir}[CIR]{Channel Impulse Response}
\acrodef{cfr}[CFR]{Channel Frequency Response}
\acrodef{csi}[CSI]{Channel State Information}
\acrodef{cs}[CS]{Compressed Sensing}
\acrodef{cv}[CV]{Constant Velocity}
\acrodef{cnn}[CNN]{Convolutional Neural Network}
\acrodef{cots}[COTS]{Commercial-Off-The-Shelf}
\acrodef{dft}[DFT]{Discrete Fourier Transform}
\acrodef{dl}[DL]{Deep Learning}
\acrodef{dma}[DMA]{Direct Memory Access}
\acrodef{dmg}[DMG]{Directional Multi Gigabit}
\acrodef{dti}[DTI]{Data Transfer Interval}
\acrodef{edmg}[EDMG]{Enhanced Directional Multi Gigabit}
\acrodef{ekf}[EKF]{Extended Kalman Filter}
\acrodef{kf}[KF]{Kalman Filter}
\acrodef{elu}[ELU]{Exponential-Linear Unit}
\acrodef{fmcw}[FMCW]{Frequency-Modulated Continuous-Wave}
\acrodef{fov}[FOV]{Field-of-View}
\acrodef{ft}[FT]{Fourier Transform}
\acrodef{fr2}[FR2]{Frequency Range 2}
\acrodef{gpio}[GPIO]{General Purpose Input/Output}
\acrodef{gsps}[GSPS]{Giga-Samples per Second}
\acrodef{har}[HAR]{Human Activity Recognition}
\acrodef{ht}[HT]{High Throughput}
\acrodef{idft}[IDFT]{Inverse Discrete Fourier Transform}
\acrodef{if}[IF]{Intermediate Frequency}
\acrodef{ifs}[IFS]{Inter-Frame Spacing}
\acrodef{iht}[IHT]{Iterative Hard Thresholding}
\acrodef{ista}[ISTA]{Iterative Shrinkage-Thresholding Algorithm}
\acrodef{isac}[ISAC]{Integrated Sensing And Communication}
\acrodef{jcs}[JCS]{Joint Communication and Sensing}
\acrodef{jpdaf}[JPDAF]{Joint Probabilistic Data Association Filter}
\acrodef{ls}[LS]{Least Squares}
\acrodef{lsd}[LSD]{Log-Spectral Distance}
\acrodef{los}[LoS]{Line-of-Sight}
\acrodef{lbm}[LBM]{Loop-Back Memory}
\acrodef{mae}[MAE]{Mean Absolute Error}
\acrodef{mcs}[MCS]{Modulation and Coding Scheme}
\acrodef{md}[$\mu$D]{micro-Doppler}
\acrodef{mimo}[MIMO]{Multiple Input Multiple Output}
\acrodef{mse}[MSE]{Mean Squared Error}
\acrodef{mmwave}[mmWave]{Millimeter-Wave}
\acrodef{msps}[MSPS]{Mega-Samples per Second}
\acrodef{mu}[MU]{Multiple User}
\acrodef{MUSIC}[MUSIC]{MUlti SIgnal Classification}
\acrodef{nac}[NAC]{Normalized Auto Correlation}
\acrodef{nco}[NCO]{Numerical Controlled Oscillator}
\acrodef{nlos}[NLoS]{Non-Line-of-Sight}
\acrodef{nn}[NN]{Neural Network}
\acrodef{nls}[NLS]{Nonlinear Least-Squares}
\acrodef{ofdm}[OFDM]{Orthogonal Frequency Division Multiplexing}
\acrodef{omp}[OMP]{Orthogonal Matching Pursuit}
\acrodef{per}[PER]{Packet Error Rate}
\acrodef{phy}[PHY]{Physical Layer}
\acrodef{pl}[PL]{Programmable Logic}
\acrodef{pov}[POV]{Point-of-View}
\acrodef{ps}[PS]{Processing System}
\acrodef{po}[PO]{Phase Offset}
\acrodef{pri}[PRI]{Pulse Repetition Interval}
\acrodef{psnr}[PSNR]{Peak Signal-to-Noise Ratio}
\acrodef{ransac}[RANSAC]{Random Sample Consensus}
\acrodef{rf}[RF]{Radio Frequency}
\acrodef{rfsoc}[RFSoC]{Radio Frequency System on a Chip}
\acrodef{rcs}[RCS]{Radar Cross-Section}
\acrodef{rss}[RSS]{Received Signal Strength}
\acrodef{rom}[ROM]{Read Only Memories}
\acrodef{rx}[RX]{receiver}
\acrodef{sc}[SC]{Single Carrier}
\acrodef{sdr}[SDR]{Software Defined Radio}
\acrodef{siso}[SISO]{Single Input Single Output}
\acrodef{sls}[SLS]{Sector Level Sweep}
\acrodef{snr}[SNR]{Signal-to-Noise Ratio}
\acrodef{ssnr}[SSNR]{Sensing Signal-to-Noise Ratio}
\acrodef{soc}[SoC]{System on a Chip}
\acrodef{spb}[SPB]{Signal Processing Blocks}
\acrodef{srrc}[SRRC]{Square-Root-Raised-Cosine}
\acrodef{ssr}[SSR]{Super Sample Rate}
\acrodef{sta}[STA]{Station}
\acrodef{std}[STD]{Standard Deviation}
\acrodef{stf}[STF]{Short Training Field}
\acrodef{stft}[STFT]{Short Time Fourier Transform}
\acrodef{su}[SU]{Single User}
\acrodef{tf}[TF]{Time-Frequency}
\acrodef{to}[TO]{Timing Offset}
\acrodef{toa}[ToA]{Time of Arrival}
\acrodef{tx}[TX]{Transmitter}
\acrodef{ula}[ULA]{Uniform Linear Array}
\acrodef{usrp}[USRP]{Universal Software Radio Peripheral}
\acrodef{vht}[VHT]{Very High Throughput}
\acrodef{wlan}[WLAN]{Wireless Local Area Network}
\acrodef{emd}[EMD]{Earth Mover's Distance}
\acrodef{6g}[6G]{6th Generation}
\acrodef{lfm}[LFM]{Linear Frequency Modulation}
\acrodef{fbmc}[FBMC]{Filter-Bank Multicarrier}
\acrodef{gfdm}[GFDM]{Generalized Frequency-Division Multiplexing}
\acrodef{dft-s-ofdm}[DFT-s-OFDM]{Discrete Fourier Transform-spread-OFDM}
\acrodef{otfs}[OTFS]{Orthogonal Time Frequency Space}
\acrodef{papr}[PAPR]{Peak-to-Average Power Ratio}
\acrodef{iot}[IoT]{Internet of Things}
\acrodef{rrss}[RRSS]{Received Raw Signal Sample}
\acrodef{csq}[CSQ]{Cellular Signal Quality}
\acrodef{pcl}[PCL]{Passive Coherent Location}
\acrodef{pslr}[PSLR]{Peak-to-Sidelobe Ratio}
\acrodef{isl}[ISL]{Integrated Sidelobe Level}
\acrodef{rd}[RD]{Range-Doppler}
\acrodef{api}[API]{Application Programming Interface}
\acrodef{agc}[AGC]{Automatic Gain Control}
\acrodef{crlb}[CRLB]{Cramér–Rao Lower Bound}
\acrodef{frft}[FRFT]{Fractional Fourier Transform}
\acrodef{osca}[OSCA]{Ordered-Statistic and Cell-Averaging}
\acrodef{cfar}[CFAR]{Constant False Alarm Rate}
\acrodef{doa}[DoA]{Direction of Arrival}
\acrodef{uav}[UAV]{Unmanned Aerial Vehicle}
\acrodef{awgn}[AWGN]{Additive White Gaussian Noise}
\acrodef{bw}[BW]{Bandwidth}
\acrodef{scs}[SCS]{Subcarrier Spacing}
\acrodef{cp}[CP]{Cyclic Prefix}
\acrodefplural{cp}[CPs]{Cyclic Prefixes}
\acrodef{pll}[PLL]{Phase-Locked Loop}

\acrodef{5g}[5G NR]{5G New Radio}
\acrodef{5ga}[5G-A]{5G-Advanced}
\acrodef{rs}[RS]{Reference Signal}
\acrodef{fr1}[FR1]{Frequency Range 1}
\acrodef{fr2}[FR2]{Frequency Range 2}
\acrodef{re}[RE]{Resource Element}
\acrodef{rg}[RG]{Resource Grid}
\acrodef{prb}[PRB]{Physical Resource Block}
\acrodef{rb}[RB]{Resource Block}
\acrodef{cran}[C-RAN]{Cloud Radio Access Network}

\acrodef{dmrs}[DM-RS]{Demodulation Reference Signal}
\acrodef{ptrs}[PT-RS]{Phase Tracking Reference Signal}
\acrodef{prs}[PRS]{Positioning Reference Signal}
\acrodef{csirs}[CSI-RS]{Channel-State Information Reference Signal}
\acrodef{pss}[PSS]{Primary Synchronization Signal}
\acrodef{sss}[SSS]{Secondary Synchronization Signal}
\acrodef{ssim}[SSIM]{Structural Similarity Index Metric}
\acrodef{ssb}[SSB]{Synchronization Signal Block}
\acrodef{rimrs}[RIM-RS]{Remote Interference Management-Reference Signal}
\acrodef{sib1}[SIB1]{System Information Block Type 1}
\acrodef{ss}[SS]{Synchronization Signal}

\acrodef{pbch}[PBCH]{Physical Broadcast Channel}
\acrodef{pdcch}[PDCCH]{Physical Downlink Control Channel}
\acrodef{pdsch}[PDSCH]{Physical Downlink Shared Channel}
\acrodef{pucch}[PUCCH]{Physical Uplink Control Channel}
\acrodef{pusch}[PUSCH]{Physical Uplink Shared Channel}

\acrodef{ue}[UE]{User Equipment}
\acrodef{gnb}[gNB]{gNodeB}

\acrodef{n1}[N1]{Node 1}
\acrodef{n2}[N2]{Node 2}
\acrodef{dbscan}[DBSCAN]{Density-Based Spatial Clustering of Applications with Noise}

\acrodef{tg}[TG]{Target}
\acrodef{rx}[RX]{Receiver}
\acrodef{rxn}[RX$_n$]{Receivers}
\acrodef{lo}[LO]{Local Oscillator}
\acrodef{t}[T]{Time}
\acrodef{path-one}[LoS]{Line of Sight}
\acrodef{path-two}[tgt]{target}
\acrodef{ifft}[IFFT]{Inverse Fast Fourier Transform}
\acrodef{mae}[MAE]{Mean Absolute Error}

\def\BibTeX{{\rm B\kern-.05em{\sc i\kern-.025em b}\kern-.08em
    T\kern-.1667em\lower.7ex\hbox{E}\kern-.125emX}}
\begin{document}

\title{Estimating Target Doppler in Unsynchronized Multistatic ISAC Deployments with Mobile Nodes\\
\thanks{* corresponding author.

This work was supported by the European Union’s Horizon 2020 research and innovation programme under the Marie Skłodowska-Curie COFUND Doctoral Programme “UNIPhD – Eight Century Legacy of Multidisciplinary Research and Training for the Next-Generation Talents”, grant agreement No. 101034319.

This publication is part of grant JDC2024-055419-I, funded by MICIU/AEI/10.13039/501100011033 and by the ESF+.

The work presented in this paper is (partially) supported by the MultiX project that has received funding from the Smart Networks and Services Joint Undertaking (SNS JU) under the European Union’s Horizon Europe research and innovation programme under Grant Agreement No 101192521.

This work has been funded by project PID2022-136769NB-I00 funded by MCIN/AEI/10.13039/501100011033 / FEDER, UE.

TUCAN6-CM (TEC-2024/COM-460), funded by Comunidad de Madrid (ORDEN 5696/2024).
}
}

\author{\IEEEauthorblockN{
Zaman Bhalli}
\IEEEauthorblockA{
\textit{University of Padova}\\
Padova, Italy \\
zaman.bhalli@unipd.it}
\and
\IEEEauthorblockN{
Michele Rossi}
\IEEEauthorblockA{
\textit{University of Padova}\\
Padova, Italy \\
michele.rossi@unipd.it}
\and
\IEEEauthorblockN{
Joerg Widmer}
\IEEEauthorblockA{
\textit{IMDEA Networks Institute}\\
Madrid, Spain \\
joerg.widmer@networks.imdea.org}
\and
\IEEEauthorblockN{
Marco Canil*}
\IEEEauthorblockA{
\textit{IMDEA Networks Institute}\\
Madrid, Spain \\
marco.canil@networks.imdea.org}
}

\maketitle

\begin{abstract}
\ac{isac} is recognized as a key enabler for future \ac{6g} networks, combining communication capabilities with pervasive sensing.
In such systems, the estimation of the Doppler shift plays a crucial role for target characterization.
However, typical real-world \ac{isac} scenarios largely involve bistatic or multistatic configurations and mobile \ac{isac} nodes. Under these conditions, Doppler estimation becomes particularly challenging, as clock asynchrony between the \ac{tx} and the \acp{rx}, combined with their mobility, introduces additional Doppler components and phase offsets that distort or disrupt the target-induced frequency shift.
Existing works have considered these challenges separately or relied on external reference reflectors.

In this paper, we present the first method to estimate the Doppler frequency of a target with mobile and asynchronous \ac{isac} nodes in a multistatic configuration, considering the case of a mobile \ac{tx} and multiple static \acp{rx}, and without leveraging any external reflector.
By leveraging the invariance of the phase offsets across multipath components and exploiting geometrical relationships, we show that the problem is solvable if at least $4$ \acp{rx} are present.
We evaluate the proposed solution through numerical simulations in various scenarios, showing that it is a valid approach for estimating target Doppler shifts in unsynchronized multistatic \ac{isac} deployments with mobile nodes.
\end{abstract}


\section{Introduction}
\label{sec:introduction}

Considered as one of the cornerstones for the \acl{6g} of wireless networks \cite{imt-2030}, \ac{isac} is expected to become an essential component of future wireless infrastructures.
Seeking to embed sensing functionalities directly into communication systems, the promise is to enable a pervasive radio-based sensing fabric that extends over the whole communication network.
In \ac{isac} applications, accurate Doppler shift estimation plays a crucial role in target characterization \cite{canil2022milliTRACE-IR} and in determining target motion \cite{pegoraro2024jump}, making it a fundamental sensing feature.
In the \ac{isac} vision, the sensing functionality will predominantly rely on bistatic configurations, i.e., with physically separated \ac{tx} and \ac{rx}, as \ac{isac} nodes are expected to consist of a combination of \acp{gnb} and \acp{ue}.
In this context, the estimation of the Doppler shift poses several additional challenges compared to monostatic cases. 
Due to the unsynchronized clocks of the \ac{tx} and \ac{rx}, the received signal is affected by \ac{to}, \ac{cfo}, and \ac{po}, which act as disturbance terms and hinder the reconstruction of the target's Doppler.
Moreover, the motion of one or both \ac{isac} nodes introduces an additional Doppler component that dynamically combines with that of the target in a non-trivial manner.

Current studies mainly address the challenges arising from clock asynchronies with \emph{static} devices \cite{pegoraro2024jump, wu2024sensing, zhao2024multiple, canil2023anexperimental}, or consider \emph{mobile} monostatic radars, which are not affected by such issues. In \cite{ventura2024bistatic} and \cite{ventura2025asymovintegratedsensingcommunications}, the authors tackle both problems at the same time, but their approach relies on the concurrent presence of \ac{los} and external static reflectors, limiting the applicability in real-world scenarios.

In this work, we present a method for estimating the Doppler of a moving target using unsynchronized and mobile \ac{isac} nodes, without relying on any external reflectors. Specifically, we consider an asynchronous multistatic configuration with one \ac{tx} and multiple \acp{rx}. Both the \ac{tx} and the target are moving, while the \acp{rx} remain static at fixed known locations.
In the envisioned scenario, the \ac{tx} corresponds to a mobile \ac{ue}, while the \acp{rx} could be different \acp{rx} of a \ac{5g} \ac{gnb} or components of a \ac{cran}, where coherent processing is possible \cite{cran_coherent}. Please, note that industrial implementations of \acp{cran} are becoming available \cite{ericsson_cloud_ran}.
An illustrative (but not limiting) example is that of a micro \ac{gnb} \cite{3gpp38901}  deployed in an urban street canyon, featuring multiple overlapping \acp{rx} with highly directional beams. Such beams can be dynamically steered and adjusted to accommodate different communication and sensing requirements.
We assume that all \acp{rx} are phase-synchronized at baseband 
but employ separate \ac{rf} front ends (i.e., with separate mixers, \ac{pll}, etc.). 
With this configuration, each \ac{rx} experiences a slightly different \ac{cfo} after down-conversion, due to the small hardware differences in the front ends. While this difference is usually negligible for communication, removing it is critical for sensing \cite{wu2024sensing}, and each \ac{rx} needs to independently compensate for its own \ac{cfo}.
In our approach, we assume that each \ac{rx} receives two paths, which appear as peaks in the estimated \ac{cir}: the \ac{los} path from the moving \ac{tx} and the target path.
\fig{geometry} represents this scenario.
The proposed methodology takes as input \emph{(i)} the phases of the received paths, \emph{(ii)} the \ac{aoa} of each path, and \emph{(iii)} the known relative position of the \acp{rx}.
Then, four steps lead to the reconstruction of the target's Doppler.
First, leveraging the fact that \ac{cfo} and \ac{po} are common to all propagation paths of the same receiver, the \ac{los} path is used to remove these offsets from the phases of the target path.
Second, range-related phase terms are removed by taking time-domain phase differences between consecutive time steps.
Third, the geometrical relationships between the phase terms and the \ac{gnb} \acp{rx} are exploited to reduce the number of independent terms. 
Finally, a non-linear system of equations is formulated and solved to estimate the target's Doppler.
As a result of the simplifications, the system becomes solvable if at least $4$ \acp{rx} are available.

The main contributions of this work are:
\begin{enumerate}
    \item We present the first method to estimate the Doppler frequency of a target in a multistatic scenario with mobile and asynchronous \ac{isac} nodes.
    \item We consider the challenging but realistic case where each \ac{isac} \ac{rx} exhibits a different \ac{cfo} and \ac{po}.
    \item We evaluate the proposed solution through numerical simulations in various scenarios, showing that it represents a valid solution for the estimation of Doppler shifts in realistic multistatic \ac{isac} scenarios.
\end{enumerate}

The remainder of the paper is organized as follows. \secref{system-modeling} and \secref{sec:methodology} describe the system model and the proposed methodology, respectively. \secref{sec:results} presents the simulation setup along with the obtained results, while \secref{sec:conclusions} provides the concluding remarks.

\section{System Description and Modeling}
\label{system-modeling}


This section presents the system architecture and model considered for our new Doppler estimation technique. 

\subsection{Reference scenario}
\label{ref-scenario}
\begin{figure}
  \centering
  \includegraphics[scale=0.125]{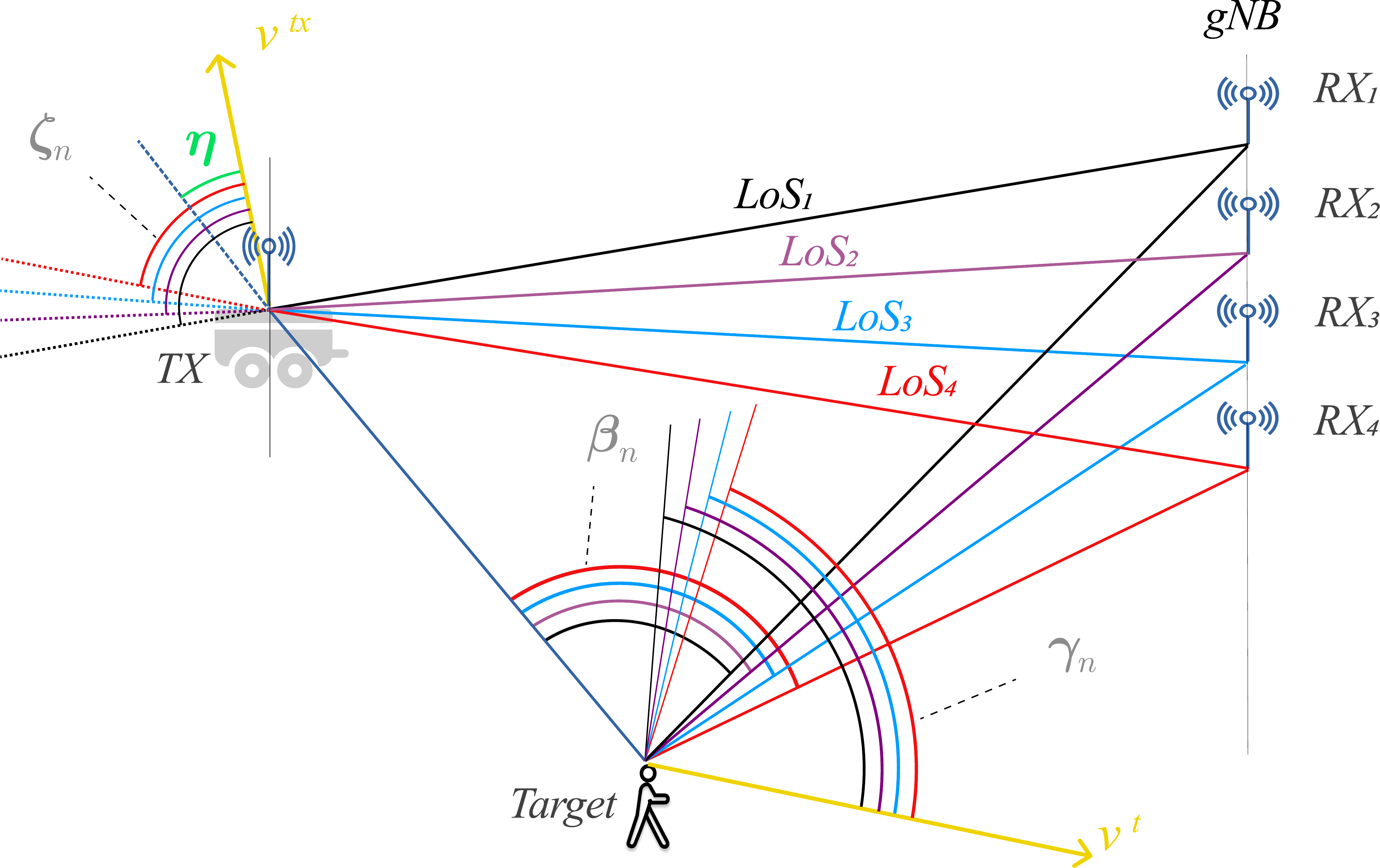}
  \caption{Geometric representation of the considered scenario with \ac{los} and target paths.}
  \label{geometry}
  \vspace{-4mm}
\end{figure}
We consider an asynchronous multistatic configuration with one \ac{tx} and multiple \acp{rx}. Both the \ac{tx} and the target are moving, while the \acp{rx}, indexed by $n=1, \dots, N_{\rm rx}$, remain static at fixed known locations.
In the envisioned scenario, the \ac{tx} corresponds to a mobile \ac{ue}, while the \acp{rx} could be different sectors of a \ac{5g} \ac{gnb} or components of a \ac{cran}, where coherent processing can be practically implemented \cite{cran_coherent, ericsson_cloud_ran}.
\fig{geometry} provides a schematic representation of the system for the case where $N_{\rm rx}=4$ and the \acp{rx} are located along a straight line with equal separation.
As represented in the figure, we assume each \ac{rx} to receive \emph{(i)} a \ac{los} path directly from the \ac{tx}, and \emph{(ii)} a target path coming from the scattering of the transmitted signal off the target. The \emph{\ac{los} path} contains contributions from the distance and the motion of the \ac{tx} and is affected by the \ac{cfo}, \ac{po}, and \ac{to} of the specific \ac{tx}-\ac{rx} pair. The \emph{target path}, instead, contains a combination of target- and \ac{tx}-induced components and is also affected by \ac{cfo}, \ac{po}, and \ac{to}.

\subsection{Continuous-Time Channel Model}
\label{sec:continuous-time}

We model the continuous-time \ac{cir} at \ac{rx} $n$, time $t$, and delay $\tau$ as \cite{proakis_dsp}:
\vspace{-2mm}
\begin{equation}
    \begin{split}
    h_n(t, \tau) = & e^{j\psi_{\mathrm{o}, n}(t)} \sum_{m=1}^{M_n(t)} A_{m,n}(t) e^{j\vartheta_{m,n}(t)} \\ 
    & \cdot \delta\bigl(\tau - \tau_{m,n}(t) - \tau_{\mathrm{o}, n}(t)\bigr) + w_n(t, \tau), \\
    \end{split}
    \label{cir-continuous}
\end{equation}
where $\delta(\cdot)$ is the Dirac delta function, $m$ index the $M_n(t)$ propagation paths, $A_{m,n}(t)$ embodies the complex path amplitude accounting for propagation loss, reflection and scattering phenomena, $\tau_{m,n}(t)$ denotes the instantaneous propagation delay, $\psi_{\mathrm{o}, n}(t)$ is the \ac{po}, $\tau_{\mathrm{o}, n}(t)$ is the \ac{to}, and $w_n(t, \tau)$ is an \ac{awgn} component. Phase term $\vartheta_{m,n}(t)$ comprises Doppler shifts and \ac{cfo} and is expressed as \cite{mimo_radar}:
\begin{equation}
    \vartheta_{m,n}(t) = 2\pi t \left(f_{\mathrm{D},m,n}^{\mathrm{tx}}(t) + f_{\mathrm{D},m,n}^{\mathrm{tgt}}(t) + f_n^{\mathrm{o}}(t)\right),
    \label{phase-components}
\end{equation}

where $f_{\mathrm{D}, m,n}^{\mathrm{tx}}(t)$, $f_{\mathrm{D},m,n}^{\mathrm{tgt}}(t)$, and $f_{n}^{\mathrm{o}}(t)$  denote the Doppler frequency of the \ac{tx}, of the target, and the \ac{cfo}, respectively.
Let us analyze the components of \eq{phase-components} for the \ac{los} and target paths. Considering the \ac{los} path, there is no Doppler shift from the target, therefore $f_{\mathrm{D},m,n}^{\mathrm{tgt}}(t)=0$. Taking \fig{geometry} as a reference and using subscripts ``LoS'' and ``tgt'' to indicate the \ac{los} and target paths, respectively, the \ac{tx} Doppler shift can be expressed as \cite{mimo_radar}:
\begin{equation}
    f_{\mathrm{D, LoS},n}^{\mathrm{tx}}(t) = \frac{\|\mathbf{v}_{\mathrm{tx}}\|}{\lambda} \cos(\zeta_n(t)),
    \label{eq:tx-dopp-@los}
\end{equation}
where $\mathbf{v}_{\mathrm{tx}}$ is the \ac{tx} velocity, $\|\cdot\|$ extracts the magnitude of a vector, $\zeta_n(t)$ is the angle between the transmitter's velocity vector and the elongation of the line connecting the \ac{rx} to the \ac{tx}, and $\lambda$ is the wavelength of the carrier wave.
Regarding the target path, the \ac{tx} Doppler is expressed as \cite{mimo_radar}: 
\begin{equation}
    f_{\mathrm{D, tgt},n}^{\mathrm{tx}}(t) = \frac{\|\mathbf{v}_{\mathrm{tx}}\|}{\lambda}\cos(\eta(t)),
    \label{eq:tx-dopp-@tgt}
\end{equation}
where $\eta(t)$ is the angle between the \ac{tx} velocity vector and the elongation of the line connecting the target to the \ac{tx}. 
The target Doppler, instead, is expressed as \cite{mimo_radar}:
\begin{equation}
    f_{\mathrm{D, tgt},n}^{\mathrm{tgt}}(t) = \frac{\|\mathbf{v}_{\mathrm{tgt}}\|}{\lambda}\cos(\gamma_{n}(t))\cos(\beta_{n}(t)/2),
    \label{eq:tgt-dopp-@tgt}
\end{equation}
where $\mathbf{v}_{\mathrm{tgt}}$ is the target's velocity vector, $\beta_n(t)$ is the angle between the \ac{tx} and the \ac{rx}, with vertex at the target, also known as the \emph{bistatic angle}, and $\gamma_{n}(t)$ is the angle between the target velocity and the bisector of the bistatic angle.
Finally, both \ac{los} and target paths are affected by the same \ac{cfo} $f_{n}^{\mathrm{o}}(t)$  which, however, only depends on the receiver $n$ and not on the path index $m$.

\subsection{Discrete-Time Measurement Model}
\label{discrete-time}
Practical implementations of \ac{isac} systems usually require the discretization of continuous-time signals. Under the common assumption that the channel's dynamic characteristics change negligibly throughout a short time interval $T$, we discretize the continuous-time \ac{cir} at \ac{rx} $n$, discrete time $k$, and delay bin $l$, as \cite{proakis_dsp}:

\begin{equation}
\begin{split}    
  h_n[k,l] = & e^{j\psi_{n}^{\mathrm{o}}[k]} \sum_{m=1}^{M_n[k]} A_{m,n}[k] e^{j\vartheta_{m,n}[k]} \\
  & \cdot \delta\left[l - {\tau}_{m,n} - \tau_{\mathrm{o,n}}[k]\right] + w_n[k,l],\\
  \end{split}
   \label{cir-discrete}
\end{equation}
where the terms are the discrete counterparts of the terms defined in \eq{cir-continuous} and the discrete version of \eq{phase-components} is \cite{proakis_dsp}:
\begin{equation}
    \vartheta_{m,n}[k] = 2\pi kT\left(f_{\mathrm{D},m,n}^{\mathrm{tx}}[k] + f_{\mathrm{D},m,n}^{\mathrm{tgt}}[k] + f_{n}^{\mathrm{o}}[k]\right).
  \end{equation}
  
  

\subsection{Measured Phase Model}
\label{sec:phase-model}
The proposed Doppler estimation technique (see \secref{sec:methodology}) primarily exploits the phases corresponding to the \ac{los} and target paths in the \ac{cir}.
We formalize the phase corresponding to the generic $m$-th path, at the $n$-th \ac{rx}, at time $k$, as:
\begin{equation}
  \label{eq:phase-model}
  \phi_{m,n}[k] = \Psi_{n}^{\mathrm{o}}[k] + \angle A_{m,n} + \chi_{m,n}[k] + \epsilon_{m,n}[k].
\end{equation}
In the formula, $\Psi_{n}^{\mathrm{o}}[k] = \psi_{n}^{\mathrm{o}}[k] + 2\pi kT f_{n}^{\mathrm{o}}[k]$ summarizes the combined effect of \ac{po} and \ac{cfo}, $\angle \cdot$ is the phase operator, $\chi_{m,n}[k] = 2\pi k T \left({f}_{\mathrm{D},m,n}^{\mathrm{tx}} +{f}_{\mathrm{D},m,n}^{\mathrm{tgt}}\right)$ captures the contribution of the Doppler components, and $\epsilon_{m,n}[k]$ is a residual noise component.
Term $\angle A_{m,n}$ indicates the phase component due to the length of the path. Considering a short-enough time window, it is reasonable to assume this value to remain constant across different time instants and we therefore drop time index $k$.

 
\section{Methodology}
\label{sec:methodology}



The methodology proposed in this paper proceeds in four main stages. First, we remove the \ac{cfo} and \ac{po} of each \ac{tx}-\ac{rx} pair, independently. Second, we perform temporal phase differentiation to eliminate the static phase terms. Third, we build a system of equations and reduce the number of unknowns by means of geometric simplifications.
Finally, we estimate the remaining unknowns, including the target's Doppler, by solving a nonlinear optimization problem.
Once the components of the Doppler frequency are estimated, we can compute the Doppler frequency of the target as a final result. 
With this procedure, we use the measurements from all the \acp{rx} to estimate the Doppler frequency of the target at \emph{one} of the \acp{rx}.
As discussed later, we found that the minimum number of \acp{rx} required for the system to be solvable is $N_{\rm rx}= 4$.

In the following, we detail each stage of the methodology.


\subsection{Removal of Carrier Frequency Offset and Phase Offset}
\label{cfo-po-removal}

As described in \secref{sec:phase-model}, $\Psi_{n}^{\mathrm{o}}[k]$ incorporates the phase terms due to \ac{cfo} and \ac{po}.
This term varies at every time step $k$, but, at a fixed $k$, is common across all multipath components received at \ac{rx} $n$.
To remove its effect, we adopt the methodology proposed in \cite{pegoraro2024jump} and we subtract the \ac{los} path from the target path at each \ac{rx}, obtaining:
\begin{equation}
  \Delta\phi_{n}[k] = \phi_{{\rm tgt},n}[k] - \phi_{{\rm LoS},n}[k],
  \label{eq:spatial-diff-phase}
\end{equation}
where subscripts ``tgt'' and ``LoS'' indicate the target and \ac{los} paths, respectively.
By substituting the right-hand terms using \eq{eq:phase-model}, \eq{eq:spatial-diff-phase} can be rewritten as:
\begin{equation}
\Delta\phi_{n}[k] = \Delta\angle A_{n} + \Delta\chi_{n}[k] + \Delta\epsilon_{n}[k],
\label{eq:spatial-diff-phase-expanded}
\end{equation}
where $\Delta\angle A_{n} = \angle A_{\text{\ac{path-two}},n} - \angle A_{\text{\ac{path-one}},n}$, $\Delta\chi_{n}[k] = \chi_{\text{\ac{path-two}},n}[k] - \chi_{\text{\ac{path-one}},n}[k]$, and $\Delta\epsilon_{n}[k]=\epsilon_{{\rm tgt}, n}-\epsilon_{{\rm LoS}, n}$.
$\Delta\chi_{n}[k]$ is explicitly expressed as:
\begin{equation}
    \Delta\chi_{n}[k] = 2\pi k T \left(f_{\mathrm{D},\text{\ac{path-two}},n}^{\mathrm{tx}}[k] +
    f_{\mathrm{D},\text{\ac{path-two}},n}^{\mathrm{tgt}}[k] -
    f_{\mathrm{D},\text{\ac{path-one}},n}^{\mathrm{tx}}[k] \right).
    \label{eq:expanded-chi}
\end{equation}


Consequently, out of the $2N_{\rm rx}$ phase measurements jointly extracted by the $N_{\rm rx}$ \acp{rx} (target and \ac{los} path for each of them), only $N_{\rm rx}$ remain after the removal of \ac{po} and \ac{cfo}.


\subsection{Removal of Static Phase Terms}
\label{static-phase-removal}

The step described in \secref{cfo-po-removal} eliminates the \ac{cfo} and \ac{po}. However, the resulting spatially differenced phases $\Delta\phi_n[k]$ still encompass the amplitude phase terms $\Delta\angle A_{n}$, which depend on the length of the paths. 
As mentioned in \secref{sec:phase-model}, if $T$ is sufficiently small, $\Delta\angle A_{n}$ can be considered constant for two subsequent time instants.
Hence, we remove it by applying a time domain phase differencing between consecutive frames.
Formally,
\begin{equation}
\Delta\phi'_{n}[k] = \phi'_{n}[k] - \phi'_{n}[k-1].
\label{temporal-diff-phase}
\end{equation}
Expanding this expression using \eq{eq:spatial-diff-phase-expanded} yields:

\begin{equation}
\Delta\phi'_{n}[k] = \Delta\chi'_{n}[k] + \Delta'\epsilon_{n}[k],
\label{eq:temporal-diff-phase-expanded}
\end{equation}
where $\Delta\chi'_{n}[k] = \Delta\chi_{n}[k] - \Delta\chi_{n}[k -1]$ and $\Delta'\epsilon_{n}[k]$ aggregates the noise resulting from the temporal differencing operation.

In \eq{eq:temporal-diff-phase-expanded}, the amplitude-related phase terms are removed, and only the Doppler-related components remain.
By using \eq{eq:expanded-chi}, $\Delta\chi'_{n}[k]$ can be formulated explicitly as:
\begin{equation}
\Delta\chi'_{n}[k] = 2\pi T \left(f_{\mathrm{D},\text{\ac{path-two}},n}^{\mathrm{tx}}[k] +
    f_{\mathrm{D},\text{\ac{path-two}},n}^{\mathrm{tgt}}[k] -
    f_{\mathrm{D},\text{\ac{path-one}},n}^{\mathrm{tx}}[k] \right).
\label{eq:temporal-doppler-phase}
\end{equation}
In \eq{eq:temporal-doppler-phase}, time difference operation reduces $kT$ to $T$.


\subsection{Geometric Simplification}
\label{sec:geometric-simplification}
In \eq{eq:temporal-doppler-phase}, the Doppler frequencies can be expanded according to \eq{eq:tx-dopp-@los}, \eq{eq:tx-dopp-@tgt}, and \eq{eq:tgt-dopp-@tgt}, resulting in:
\begin{equation}
\begin{split}
\Delta\chi'_{n} 
= & 2\pi T \Bigg( \frac{\|\vv{v}_{\rm tx}\|}{\lambda} \cos(\eta ) + \\ 
& + \frac{\|\vv{v}_{\rm tgt}\|}{\lambda} \, \cos(\gamma_{n})\cos\left(\frac{\beta_{n}}{2}\right) + \\
& - \frac{\|\vv{v}_{\rm tx}\|}{\lambda} \cos(\zeta_n)\Bigg),
\end{split}
\label{eq:bistatic-doppler-expanded}
\end{equation}
with $n=1, \dots, N_{\rm rx}$ and where the time index $k$ has been dropped to improve clarity. 
This set of equations forms a system of $N_{\rm rx}$ equations in $3N_{\rm rx}+3$ unknowns, namely $\vv{v}_{\rm tx}, \vv{v}_{\rm tgt}, \eta, \gamma_n, \beta_n,$ and $\zeta_n$.
$\Delta\chi'_n$, on the left-hand side, can be computed based on the measured phases.

In this shape, the system results underdetermined and cannot be solved.
Therefore, in the following, we derive some geometrical simplifications that allow to reduce the number of unknowns of the system (or, equivalently, to increase the number of independent equations).
In particular, the simplifications are based on the known \acp{aoa} of the received paths and on the the known relative positions of the \acp{rx}.

\begin{figure}[t]
  \centering
  \includegraphics[scale=0.115]{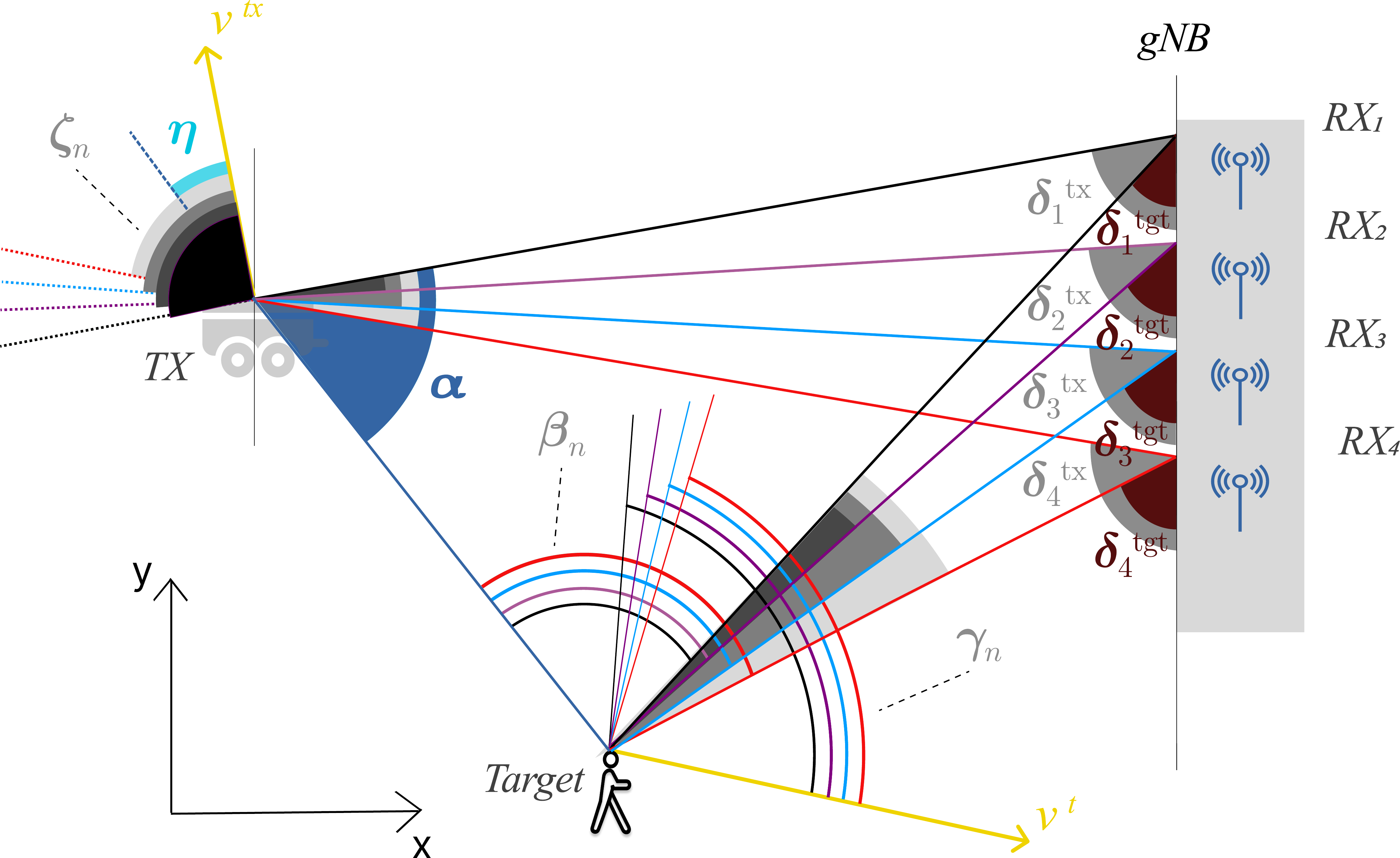}
  \caption{Representation of the paths and angles.}
  \label{fig:simplified-geometry}
  \vspace{-4mm}
\end{figure}

For simplicity, let us consider the scenario presented in \fig{fig:simplified-geometry}, where $N_{\rm rx}=4$ and the \acp{rx} are placed along a vertical line, and let us take the \ac{rx} with $n=1$ as a reference \ac{rx} for the subsequent derivations.
The computations can be easily extended to arbitrary geometries.
By combining the known locations of the \acp{rx} and the \acp{aoa} of the received paths, we use the $n$ \ac{los} paths to localize the \ac{tx}, and the $n$ target paths to localize the target. 
For each pair of \acp{rx}, an estimate of the target and \ac{tx} locations is computed by finding the intersection point of the lines originating from the \acp{rx} and following the direction of the respective \acp{aoa}.
To improve robustness, we compute such estimates for all the possible combinations of \acp{rx} pairs, and then we take their average as the final estimate.
Then, the estimated locations of target and \ac{tx} are used to build $N_{\rm rx}$ triangles with vertices at \ac{tx}, the target, and the $n$-th \ac{rx}, from which angles $\beta_n$ and $\alpha$ are estimated. 
Angle $\alpha$ is defined as in \fig{fig:simplified-geometry}.
Then, we note that each angle $\zeta_{n+1}$ and $\gamma_{n+1}$ can be written as a function of the preceding angles $\zeta_{n}$, $\gamma_{n}$, respectively, as illustrated in \fig{fig:simplified-geometry}.
Formally,
\begin{equation}
\label{eq:angle-relations}
\left\{
\begin{aligned}
    \zeta_{n+1} &= \zeta_n - (\delta^{\rm tx}_{n+1} - \delta^{\rm tx}_{n}) \\
    \gamma_{n+1} &= \gamma_n - (\delta^{\rm tx}_{n+1} - \delta^{\rm tx}_{n})/2 \\
\end{aligned}
\right.
\end{equation}
for $n=1, \dots, N_{\rm rx} - 1$.
Here, $\delta^{\rm tx}_n$ and $\delta^{\rm tgt}_n$ represent the \ac{aoa} of the \ac{los} and target path at the $n$-th \ac{rx}, respectively. Note that, in \eq{eq:angle-relations}, the only unknowns are $\zeta_1$, $\gamma_1$, as the others can be recursively derived from them.
We also note that
\begin{equation}
    \eta = \zeta_1 - \alpha.
    \label{eq:eta}
\end{equation}
Each equation in \eq{eq:angle-relations} allows for the reduction of $N_{\rm rx}-1$ unknowns, \eq{eq:eta} allows to remove $1$ more unknown ($\eta$), and the values of all the $N_{\rm rx}$ angles $\beta_n$ were previously found through the construction of the triangles. Therefore, the system of $N_{\rm rx}$ equations in $3N_{\rm rx}+3$ unknowns of \eq{eq:bistatic-doppler-expanded} is now reduced to a system with $N_{\rm rx}$ equations but $3N_{\rm rx}+3 - [2(N_{\rm rx}-1)+1+N_{\rm rx}]=4$ unknowns, and is solvable when $N_{\rm rx}\ge 4$.
\subsection{System Solution and Scalability Considerations}
After the abovementioned simplifications, the system of equations in \eq{eq:bistatic-doppler-expanded} only contains $4$ unknowns, namely $\vv{v}_{\rm tx}, \vv{v}_{\rm tgt}, \zeta_1$, and $\gamma_1$.
To find them, we build a nonlinear \ac{ls} problem and we solve it through the Levenberg–Marquardt algorithm.
To improve the quality of the estimation, after solving the system, we compute a moving average of the previous $N_w$ solutions and use that as the final estimate.
Finally, the target's Doppler frequency at one of the \acp{rx} is computed through \eq{eq:tgt-dopp-@tgt}.
Thanks to the properties of \ac{ls} 
and to the presented simplifications, the proposed solution directly scales to cases with more than $4$ \acp{rx}, as each new \ac{rx} provides an additional equation to the system, but no additional unknowns (see the end of \secref{sec:geometric-simplification}), thus only increasing its statistical redundancy.
If $N_{\rm rx} <4$, the system becomes underdetermined and, therefore, does not have a unique solution. However, if one or more \acp{rx} are \emph{temporarily} unavailable for a short period of time, tracking and filtering techniques can help to compensate for the missing equations.




\section{Numerical Results and Analysis}
\label{sec:results}
In this section, we describe the simulation setup and the obtained results.

\subsection{Simulation Setup}
To evaluate the performance of the proposed methodology, we generate standard-compliant \ac{5g} \ac{pusch} \acp{dmrs} using the \textsc{5G Toolbox} of \textsc{matlab} \cite{matlab5gtoolbox2023b} at carrier frequency $f_c=28$~GHz and we use them as transmitted signals in a simulator that we developed in Python.
We transmit 1 \ac{dmrs} per slot.
In the simulator, each target is represented by a point moving in the space with a \ac{rcs} $\rho\sim\textrm{Lognormal}(\mu_{\rm rcs}, \sigma_{\rm rcs}^2)$ that follows a log-normal distribution, where $\mu_{\rm rcs}$ and $\sigma_{\rm rcs}$ are the mean and variance of $\log(\rho)$, to account for the \ac{rcs} oscillations. 
The channel follows an \ac{awgn} model with noise spectral density of $-174$~dB/Hz (corresponding to a thermal noise of $290$~K, and the \ac{tx} uses a power of $200$~mW, corresponding to the maximum allowed for a \ac{ue} uplink according to \ac{3gpp} specifications \cite{3gpp38101-1}. 
At the  \ac{rx} side, a least-squares channel estimation is performed \cite{beek1995onchannel} using the received \ac{dmrs} symbols, and the subcarriers with no \ac{dmrs} are filled via linear interpolation. Then, \ac{cfo} and \ac{po} are added independently to each \ac{rx}, corresponding to the case where the \acp{rx} are phase-synchronized at baseband 
but have different front ends.
From the \ac{cir} estimated at each \ac{rx}, the phases of the \ac{los} and target peaks are extracted and fed to our algorithm, along with the corresponding \acp{aoa}, whose error is modeled as an additive Gaussian noise $\mathcal{N}(0, \sigma_{\delta}^2)$.
\tab{tab:simulation_parameters} summarizes the simulation parameters.


\begin{table}[H]
	\centering
    \begin{tabular}{lcc}
        \toprule
        Parameter & Symbol & Value \\
        \midrule
        Carrier frequency & $f_c$ & $28$~GHz \\
        Subcarrier spacing & SCS & $120$~kHz \\
        Number of resource blocks & $N_{\rm RB}$ & 52 \\
        Time slot duration & $T_{\rm slot}$ & $125\mu$s \\
        \ac{pusch} mapping type & / & A \\
        \ac{dmrs} configuration type & / & 2 \\
        \ac{tx} power & $P_{\rm tx}$ & $200$~mW \\
        Noise spectral density & $N_0$ & $-174$~dB/Hz \\
        \ac{rcs} mean parameter & $\mu{\rm rcs}$ & 50 \\
        \ac{rcs} variance parameter & $\sigma_{\rm rcs}^2$ & 100 \\
        \bottomrule
    \end{tabular}
	\caption{Simulation parameters.}
    \label{tab:simulation_parameters}
\end{table}

\subsection{Simulation Results}
We evaluate the performance of our method by computing the \ac{mae} of the reconstructed Doppler as $MAE= \frac{1}{K} \sum\limits_{k=1}^{K} \left| f_{\rm D}[k] - \hat{f}_{\rm D}[k] \right|$, where $f_{\rm D}[k]$ and $\hat{f}_{\rm D}[k]$ represent the true and reconstructed Doppler, respectively, and $K$ is the total number of time steps.
In the boxplots, the mean and the median are represented by a dashed and a solid line, respectively.
In the evaluation, we reconstruct the Doppler at \ac{rx} $n=1$, but any \ac{rx} could be equivalently selected.
We remind that, in order to reconstruct the Doppler at one \ac{rx}, we need to have phase and \ac{aoa} data from $N_{\rm rx}\ge4$ receivers.

We consider a reference geometry for the simulations and then vary some of the parameters to assess their impact.
The considered scenario is meant to represent a placement of the \acp{rx} loosely compatible with that of a \ac{gnb} with multiple sectors, with the different \acp{rx} corresponding to the sectors, and the \acp{rx} deployed in a localized area.
According to the axes represented in \fig{fig:simplified-geometry}, we place the \ac{tx}, target, and \acp{rx} as described in \tab{tab:reference_scenario_params}. We indicate with $\text{RX}_n$ the $n$-th \ac{rx}.
The velocity angle is computed counterclockwise with respect to the $x$ axis.
For the moving average, we take $N_w=500$.

For each set of parameters, we performed $12$ simulations.
\vspace{-2mm}


\vspace{0.1in}
\begin{table}[H]
    \centering
    \begin{tabular}{l c l c}
        \toprule
        Parameter & Value & Parameter & Value \\
        \midrule
        \ac{tx} position           & $[8,\;0]$                & \ac{aoa} noise STD         & $2^{\circ}$ \\
        Target position            & $[12,\;-12]$             & \ac{tx} velocity            & $4~\mathrm{m/s}$ \\
        $\ac{rx}_1$ position       & $[15,\;4]$               & Target velocity             & $4~\mathrm{m/s}$ \\
        $\ac{rx}_2$ position       & $[15,\;2]$               & \ac{tx} velocity angle      & $225^{\circ}$ \\
        $\ac{rx}_3$ position       & $[15,\;0]$               & Target velocity angle       & $315^{\circ}$ \\
        $\ac{rx}_4$ position       & $[15,\;-2]$              & \\
        \bottomrule     
    \end{tabular}
    \caption{Reference scenario parameters.}
    \label{tab:reference_scenario_params}
\end{table}

\begin{figure}[t]
    \centering
    \begin{subfigure}{0.48\columnwidth}
        \centering
        \includegraphics[width=\linewidth]{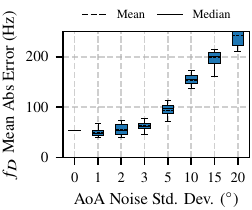}
        \caption{Varying \ac{aoa} noise.}
        \label{Varying_aoa_noise}
    \end{subfigure}
       \begin{subfigure}{0.48\columnwidth}
        \centering
        \includegraphics[width=\linewidth]{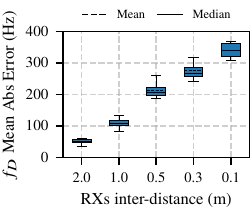}
        \caption{Varying \acp{rx} inter-distance.}
        \label{Varying_rx_dist}
    \end{subfigure}
    \caption{ Target Doppler \ac{mae} versus (a) \ac{aoa} noise and (b) RXs inter-distance}
    \label{fig: Varying_tx_pos and Varying_aoa_noise}
    \vspace{-3mm}
\end{figure}

\subsubsection{Impact of AoA noise}
We start the evaluation by exploring the impact of noise on the \acp{aoa}. This is a critical aspect of the proposed solution, since many derivations rely on \ac{aoa} values.
We vary the \ac{aoa} noise \ac{std} as 
presented in \fig{Varying_aoa_noise}. 
As expected, the Doppler reconstruction error increases with $\sigma_{\delta}$, with small $\sigma_{\delta}$ values (1–3$^{\circ}$) causing a modest degradation, and larger ones (10–20$^{\circ}$) significantly impacting the results.
We consider these results satisfactory, as $MAE\le100$~Hz for $\sigma_{\delta}\le3^\circ$ and this level of \ac{aoa} estimation accuracy is easily achievable by a \ac{rx} array with $4-8$ antennas or more \cite{mimo_radar}.
We select $\sigma_{\delta}=2^\circ$ as a realistic value and adopt it in the following evaluations.

\subsubsection{Impact of RXs inter-distance}
We study how the distance between the receivers affects the target Doppler estimate, as presented in \fig{Varying_rx_dist}.
The results show that a larger separation between the \acp{rx}, that is, a higher spatial diversity, provides better performance.
As the inter-distance decreases, the paths become more similar, making it more difficult to understand the scene dynamics.
Since the results drop significantly as the \acp{rx} inter-distance decreases, we pick the largest distance of $2$~m and we keep it for the subsequent simulations.

\subsubsection{Impact of TX and target distance}
We explore the impact of the TX/target distance from the \acp{rx}, which allows us to characterize the operating range of the system.
For the \ac{tx}, we set the starting location at $y=0$~m and we vary $x$ as $\{10, 7, 5, 0\}$~m, corresponding to a distance from the \acp{rx} of $\{5, 7, 10, 15\}$~m, respectively.
For the target, instead, we consider two starting locations: one that is near the \acp{rx}, labeled as ``TGT NEAR'', and one that is far from the \acp{rx}, labeled as ``TGT FAR''.
The near location is set at $(12, -12)$, while the far location is set at $(8, -12)$.
Since, in our scenario, we expect the distance along the $x$ axis to be more significant than that along the $y$ axis, the terms ``near'' and ``far'' refer to the distance along the $x$ axis.
\fig{Varying_tx_pos} shows the results. For this particular case, to better visualize the wide dynamic range of the $MAE$, the logarithmic scale on the $y$ axis is used.
We observe that, in the ``near'' case, the results are satisfactory up to $10$~m of distance of the \ac{tx}, while in the ``far'' case, they are generally worse.
We conclude that, with the considered placement of the \acp{rx}, the method works well only at short range, at distances compatible with those of, e.g., a narrow street.
We briefly extend this examination by considering a different geometry for the placement of the \acp{rx}, to verify if and how a higher spatial diversity helps in dealing with longer distances.
We place \ac{rx}$_1$, \ac{rx}$_2$, \ac{rx}$_3$, and \ac{rx}$_4$ at locations $(15, -2)$, $(15, 4)$, $(-15, 4)$, and $(-15, -2)$, respectively. Basically, we keep two \acp{rx} at one side and we locate the other two at the other side, symmetrically with respect to $x=0$, corresponding, e.g., to a deployment in a $30$~m wide large street with $2$ \acp{rx} per side.
Then, we consider the worst case of \fig{Varying_tx_pos}, that is, when the distance of the \ac{tx} is $15$~m, and we evaluate it with the new configuration.
The results are labeled in \fig{Varying_tx_pos} as ``15nc''.
As we can see, with higher spatial diversity, even the worst case scenario from before is tackled effectively.
The downside of this geometrical configuration is that a distributed deployment of coherent \acp{rx} is much more challenging than a localized one, as that taken as a reference in this paper.
\subsubsection{Impact of target speed}
\begin{figure}[t]
    \centering
     \begin{subfigure}{0.48\columnwidth}
        \centering
        \includegraphics[width=\linewidth]{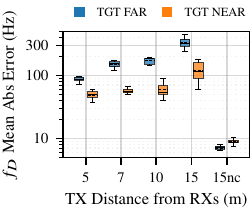}
        \caption{Varying \ac{tx}/target position.}
        \label{Varying_tx_pos}
    \end{subfigure}
    \begin{subfigure}{0.48\columnwidth}
        \centering
        \includegraphics[width=\linewidth]{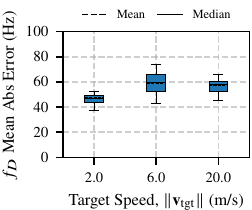}
        \caption{Varying target speed.}
        \label{Varying_tgt_speed}
    \end{subfigure}
    \caption{\ac{mae} of the target Doppler as a function of (a) distance from the \acp{rx} and (b) target speed.}
    \label{varying speed and inter-distance rxs}
    \vspace{-3mm}
\end{figure}

To study the impact of the target speed we consider $3$ velocities, $\{2, 6, 20\}$~m/s, representative of the typical speeds of a pedestrian, a bicycle, and a car, respectively.
Since we assessed that the distance along the $x$ axis has a strong impact on the results, to isolate the effect of the speed only, for these simulations we set the angle of the velocities to $270^\circ$, that is, towards the negative direction of the $y$ axis.
The results, presented in \fig{Varying_tgt_speed}, show that our method is invariant to the speed of the target.



\section{Conclusions}
\label{sec:conclusions}
In this paper, we dealt with the problem of estimating the Doppler shift of a target in a multistatic scenario with a mobile \ac{tx} and multiple \acp{rx}, showing that, with at least $4$ \ac{rx}, the problem is solvable. Simulation results show that with a localized placement of the \acp{rx} (e.g., \ac{gnb}-like), the proposed methodology works well at short range.
As briefly discussed, a distributed placement of the \acp{rx} could, instead, extend the solution operating range, at the cost of an increased implementation complexity.
Future works involve the optimization of the placement of the \acp{rx} and validating experimentally the proposed methodology.

\bibliography{refs.bib}

@article{pegoraro2024jump,
  title={JUMP: Joint communication and sensing with Unsynchronized transceivers Made Practical},
  author={Pegoraro, Jacopo and Lacruz, Jesus O and Azzino, Tommy and Mezzavilla, Marco and Rossi, Michele and Widmer, Joerg and Rangan, Sundeep},
  journal={IEEE Transactions on Wireless Communications},
  year={2024},
  publisher={IEEE}
}

@article{wu2024sensing,
  title={Sensing in Bistatic ISAC Systems With Clock Asynchronism: A signal processing perspective [Special Issue on Signal Processing for the Integrated Sensing and Communications Revolution]},
  author={Wu, Kai and Pegoraro, Jacopo and Meneghello, Francesca and Zhang, J Andrew and Lacruz, Jesus O and Widmer, Joerg and Restuccia, Francesco and Rossi, Michele and Huang, Xiaojing and Zhang, Daqing and others},
  journal={IEEE Signal Processing Magazine},
  volume={41},
  number={5},
  pages={31--43},
  year={2024},
  publisher={IEEE}
}

@article{ventura2024bistatic,
  title={Bistatic Doppler Frequency Estimation with Asynchronous Moving Devices for Integrated Sensing and Communications},
  author={Ventura, Gianmaria and Bhalli, Zaman and Rossi, Michele and Pegoraro, Jacopo},
  journal={IEEE Wireless Communications Letters},
  year={2024},
  publisher={IEEE}
}

@misc{matlab5gtoolbox2023b,
  author       = {{The MathWorks, Inc.}},
  title        = {{5G Toolbox (R2023b)}},
  year         = {2023},
  address      = {Natick, Massachusetts, United States},
  howpublished = {[Computer software]},
  url          = {https://www.mathworks.com/products/5g.html}
}

@misc{ericsson_cloud_ran,
  author = {Ericsson},
  title = {{Ericsson Cloud RAN}},
  url = {https://www.ericsson.com/en/ran/cloud},
  year = {2025},
  note = {Accessed: 2026-01-20},
  address = {Stockholm, Sweden}
}

@misc{3gpp38101-1,
   author={3rd Generation Partnership Project}, 
   title={{3GPP TS 38.101-1 version 18.6.0 Release 17}}, 
   number = "TS 38.101-1 v18.6.0",
   year= {2024},
   url = {https://www.etsi.org/deliver/etsi_ts/138100_138199/13810101/18.06.00_60/ts_13810101v180600p.pdf}
}

@misc{3gpp38901,
   author={3rd Generation Partnership Project}, 
   title={{3GPP TR 38.901 version 16.1.0 Release 16}}, 
   number = "TR 38.901 v16.1",
   year= {2020},
   url = {https://www.etsi.org/deliver/etsi_tr/138900_138999/138901/16.01.00_60/tr_138901v160100p.pdf}
}

@misc{imt-2030,
  author = {{ITU}},
  title = {{Framework and overall objectives of the future development of IMT for 2030 (6G) and beyond}},
  number = {M.2160},
  url = {https://www.itu.int/rec/R-REC-M.2160-0-202311-I/en},
  year = {2023},
}

@INPROCEEDINGS{beek1995onchannel,
  author={van de Beek, J.-J. and Edfors, O. and Sandell, M. and Wilson, S.K. and Borjesson, P.O.},
  booktitle={1995 IEEE 45th Vehicular Technology Conference. Countdown to the Wireless Twenty-First Century}, 
  title={On channel estimation in OFDM systems}, 
  year={1995},
  volume={2},
  number={},
  pages={815-819 vol.2},
  doi={10.1109/VETEC.1995.504981}}

@ARTICLE{zhao2024multiple,
  author={Zhao, Jingbo and Lu, Zhaoming and Zhang, J. Andrew and Dong, Shixu and Zhou, Shuang},
  journal={IEEE Transactions on Vehicular Technology}, 
  title={Multiple-Target Doppler Frequency Estimation in ISAC With Clock Asynchronism}, 
  year={2024},
  volume={73},
  number={1},
  pages={1382-1387},
  doi={10.1109/TVT.2023.3301589}}

@inproceedings{canil2023anexperimental,
author = {Canil, Marco and Pegoraro, Jacopo and Lacruz, Jesus O. and Mezzavilla, Marco and Rossi, Michele and Widmer, Joerg and Rangan, Sundeep},
title = {An Experimental Prototype for Multistatic Asynchronous ISAC},
year = {2023},
isbn = {9798400704413},
publisher = {Association for Computing Machinery},
address = {New York, NY, USA},
url = {https://doi.org/10.1145/3628357.3629710},
doi = {10.1145/3628357.3629710},
booktitle = {Proceedings of the First ACM Workshop on MmWave Sensing Systems and Applications},
pages = {16–17},
numpages = {2},
location = {Istanbul, Turkiye},
series = {mmWave '23}
}

@misc{ventura2025asymovintegratedsensingcommunications,
      title={AsyMov: Integrated Sensing and Communications with Asynchronous Moving Devices}, 
      author={Gianmaria Ventura and Michele Rossi and Jacopo Pegoraro},
      year={2025},
      eprint={2412.10387},
      archivePrefix={arXiv},
      primaryClass={eess.SP},
      url={https://arxiv.org/abs/2412.10387}, 
}

@ARTICLE{canil2022milliTRACE-IR,
  author={Canil, Marco and Pegoraro, Jacopo and Rossi, Michele},
  journal={IEEE Journal of Selected Topics in Signal Processing}, 
  title={milliTRACE-IR: Contact Tracing and Temperature Screening via mmWave and Infrared Sensing}, 
  year={2022},
  volume={16},
  number={2},
  pages={208-223},
  doi={10.1109/JSTSP.2021.3138632}}

@article{cran_coherent, title={{Phase and Time Synchronization for 5G C-RAN: Requirements, Design Challenges and Recent Advances in Standardization}}, volume={5}, url={https://publications.eai.eu/index.php/inis/article/view/401}, DOI={10.4108/eai.27-6-2018.155238}, number={15}, journal={EAI Endorsed Transactions on Industrial Networks and Intelligent Systems}, author={Philip Venmani, Daniel and Lagadec, Yannick and Lemoult, Olivier and Deletre, Fabrice}, year={2018}, month={Aug.}, pages={e3} }

@book{mimo_radar,
  title={MIMO radar signal processing},
  author={Li, Jian and Stoica, Petre},
  year={2008},
  publisher={John Wiley \& Sons}
}

@book{proakis_dsp,
author = {Proakis, John G. and Manolakis, Dimitris G.},
title = {Digital signal processing (3rd ed.): principles, algorithms, and applications},
year = {1996},
isbn = {0133737624},
publisher = {Prentice-Hall, Inc.},
address = {USA}
}
\bibliographystyle{IEEEtran}
\end{document}